# THE NEW MODIFIED VLASOV EQUATION FOR THE SYSTEMS WITH DISSIPATIVE PROCESSES

**E.E. Perepelkin[a], B.I. Sadovnikov[a], N.G. Inozemtseva[b]**

[a] *Faculty of Physics, Lomonosov Moscow State University, Moscow, 119991 Russia*
*E. Perepelkin e-mail: pevgeny@jinr.ru, B. Sadovnikov sadovnikov@phys.msu.ru*
[b] *Dubna State University, Moscow region, Moscow,141980 Russia*
*e-mail: nginozv@mail.ru*

**Abstract**
A new modified Vlasov equation has been obtained in this paper for the systems with dissipative phenomena such as, for example, plasma with irradiation.

**Key words:** rigorous result, Vlasov equation, entropy, plasma, thermonuclear fusion, accelerator physics

**Introduction**

Ostrogradsky [1] considered in 1850 physical systems with high order kinematic characteristics $\vec{r}, \vec{v}, \dot{\vec{v}}, \ddot{\vec{v}}, ...$ (coordinate, velocity, acceleration and high order acceleration). This approach is known in the literature as high order mechanics or Ostrogradsky mechanics [2, 3]. The quantum field theory, the gauge theories [4-6], general relativity [7-12] and some string models [13, 14] use the high order mechanics.

In the middle of the 20th century, Vlasov considered in [15-18] the similar method. He considered the probability distribution density functions depending not only on coordinate $\vec{r}$ and velocity $\vec{v}$ but also on acceleration and high order acceleration:

$$\begin{aligned} &f_1(\vec{r},t), \\ &f_2(\vec{r},\vec{v},t), \\ &f_3(\vec{r},\vec{v},\dot{\vec{v}},t), \\ &... \end{aligned} \quad (i.1)$$

where $\vec{r}, \vec{v}, \dot{\vec{v}}, \ddot{\vec{v}}, ...$ are independent variables.

Probability density functions (i.1) satisfy the relations [15-21]

$$f_0(t) \stackrel{det}{=} N(t) = \int_{(\infty)} f_1(\vec{r},t) d^3r = \int_{(\infty)}\int_{(\infty)} f_2(\vec{r},\vec{v},t) d^3r d^3v = ... =$$
$$= ... = \int_{(\infty)}\int_{(\infty)}\int_{(\infty)} ... f(\vec{r},\vec{v},\dot{\vec{v}},...,t) d^3r d^3v d^3\dot{v}... \quad , \quad (i.2)$$

where $N(t)$ is normalization factor or a number of particles, which can be non-integer in general case [15-21].

Vlasov in [15, 16] obtained the infinite chain of equations for the probability distribution density functions (i.1):

$$\frac{\partial f_1(\vec{r},t)}{\partial t} + \mathrm{div}_r \int\limits_{(\infty)} f_2(\vec{r},\vec{v},t)\vec{v}\,d^3v = 0,$$

$$\frac{\partial f_2(\vec{r},\vec{v},t)}{\partial t} + \mathrm{div}_r\left[f_2(\vec{r},\vec{v},t)\vec{v}\right] + \mathrm{div}_v \int\limits_{(\infty)} f_3(\vec{r},\vec{v},\dot{\vec{v}},t)\dot{\vec{v}}\,d^3\dot{v} = 0, \qquad (i.3)$$

...

$$\frac{\partial f_n\!\left(\vec{r},\vec{v},...,\overset{(n-1)}{\vec{r}},t\right)}{\partial t} + \mathrm{div}_r\left[f_n\!\left(\vec{r},\vec{v},...,\overset{(n-1)}{\vec{r}},t\right)\vec{v}\right] + ... + \mathrm{div}_{\overset{(n-1)}{r}} \int\limits_{(\infty)} f_{n+1}\!\left(\vec{r},\vec{v},...,\overset{(n)}{\vec{r}},t\right)\overset{(n)}{\vec{r}}\,d^3\overset{(n)}{\vec{r}} = 0,$$

...

The first equation from the Vlasov equation chain (i.3) may be written in the form of the continuity equation for the probability distribution density function $f_1(\vec{r},t)$:

$$\frac{\partial f_1(\vec{r},t)}{\partial t} + \mathrm{div}_r\left[f_1(\vec{r},t)\langle\vec{v}\rangle(\vec{r},t)\right] = 0, \qquad (i.4)$$

where by definition

$$f_1(\vec{r},t)\langle\vec{v}\rangle(\vec{r},t) \overset{\mathrm{det}}{=} \int\limits_{(\infty)} f_2(\vec{r},\vec{v},t)\vec{v}\,d^3v.$$

Paper [22] described the properties of equation (i.4) and showed its link with the Schrödinger equation by using substitution $f_1(\vec{r},t) = |\Psi(\vec{r},t)|^2$.

The second equation from the Vlasov equation chain (i.3) can be rewritten in the form:

$$\frac{\partial f_2(\vec{r},\vec{v},t)}{\partial t} + \mathrm{div}_r\left[f_2(\vec{r},\vec{v},t)\vec{v}\right] + \mathrm{div}_v\left[f_2(\vec{r},\vec{v},t)\langle\dot{\vec{v}}\rangle(\vec{r},\vec{v},t)\right] = 0, \qquad (i.5)$$

where by definition

$$f_2(\vec{r},\vec{v},t)\langle\dot{\vec{v}}\rangle(\vec{r},\vec{v},t) \overset{\mathrm{det}}{=} \int\limits_{(\infty)} f_3(\vec{r},\vec{v},\dot{\vec{v}},t)\dot{\vec{v}}\,d^3\dot{v}.$$

Vlasov rewrote equation (i.5) in the form

$$\frac{\partial f_2}{\partial t} + f_2\,\mathrm{div}_r[\vec{v}] + (\vec{v},\nabla_r f_2) + f_2\,\mathrm{div}_v\langle\dot{\vec{v}}\rangle + (\langle\dot{\vec{v}}\rangle,\nabla_v f_2) = 0. \qquad (i.6)$$

As $\vec{r}$ and $\vec{v}$ are independent variables, then $\mathrm{div}_r[\vec{v}] = 0$ and equation (i.6) is of the form

$$\frac{\partial f}{\partial t} + (\vec{v},\nabla_r f) + (\langle\dot{\vec{v}}\rangle,\nabla_v f) + f\,\mathrm{div}_v\langle\dot{\vec{v}}\rangle = 0. \qquad (i.7)$$

For equation (i.7) Vlasov considered the following particular case:

$$\begin{aligned}&1. \quad \langle \dot{\vec{v}} \rangle = \frac{d}{dt}\langle \vec{v} \rangle = \frac{\vec{F}}{m}, \\ &2. \quad \text{div}_v \langle \dot{\vec{v}} \rangle = 0.\end{aligned} \qquad (i.8)$$

Substituting expression (i.8) into equation (i.7) gives the famous Vlasov equation

$$\frac{\partial f}{\partial t} + (\vec{v}, \nabla_r f) + \left(\frac{\vec{F}}{m}, \nabla_v f\right) = 0, \qquad (i.9)$$

where the substitution of $\vec{F} = q\left(\vec{E} + [\vec{v}, \vec{B}]\right)$ into equation (i.9) gives the Vlasov-Poisson equation in the particular case.

A large number of articles on statistical mechanics, plasma physics, thermonuclear fusion, accelerator physics, astrophysics and condense matter physics considers the classical Vlasov equation (i.9).

Note there are many different derivations of the Vlasov equation (i.9) today. However, following the chronological order, Vlasov himself obtained equation (i.9) from equation chain (i.3) by cutting it off at the second equation (i.5) [15, 16].

Equation (i.9) is equal to the Liouville's equation. As a result, a broad class of dissipative systems and systems with non-constant entropy cannot be described by equation (i.9). For example, the electromagnetic irradiation power depends on $\dot{v}^2$, but due to conditions (i.8) equation (i.9) is unsuitable for description of such systems.

***The goal of this work is to consider equation (i.5) without applying conditions (i.8).***

The main definitions for the average values of velocities, accelerations and high order accelerations are given in §1. The concept of the entropy of a continuous random value, which is characterized by high order probability density functions (i.1), is defined as well. Differential operator $\Pi_n$ is introduced in §1 for the compact form of the infinite Vlasov equation chain. The properties of operator $\Pi_n$ is considered in §3.

The equation chain for entropy has been obtained in §2. The equation chain for entropy corresponds to the Vlasov equation chain (i.3). From the equation chain for entropy, it follows that the changes of the system entropy are defined by $\text{div}_r \langle \vec{v} \rangle, \text{div}_v \langle \dot{\vec{v}} \rangle, \text{div}_{\dot{v}} \langle \ddot{\vec{v}} \rangle, \ldots$ values, i.e. by existence of "sources" of high order kinematical vector fields. It has been shown that without such "sources" the system entropy is constant.

In §3,4 the general expression for $\langle \dot{\vec{v}} \rangle$ versus $\frac{d}{dt}\langle \vec{v} \rangle$, for $\langle \langle \ddot{\vec{v}} \rangle \rangle$ versus $\frac{d}{dt}\langle \dot{\vec{v}} \rangle$ and so on has been obtained. In particular, it has been shown that conditions (i.8) are not fulfilled in general case, i.e. $m\langle \dot{\vec{v}} \rangle \neq m\frac{d}{dt}\langle \vec{v} \rangle = \vec{F}$. The new modified Vlasov equation for the general case of non-constant entropy has been obtained in §4. The new modified Vlasov equation can be used for the dissipative system description.

The main lemmas and theorems are stated in Paragraphs 1-4 and its proofs are placed in Section «Appendices» for convenience. Conclusion presents the obtained results.

### 1. Differential operator $\Pi_n$

The infinite chain of the Vlasov equations for the probability density functions $f_n(\vec{\xi}_n, t) = f_n(\vec{r}, \vec{v}, \dot{\vec{v}}, ..., t)$, $n \in \mathbb{N}$, $\vec{\xi}_n \in \Omega$, (where $\Omega$ − generalized phase space [19-21]) is of the following form [15,16]:

$$\frac{\partial f_1(\vec{r},t)}{\partial t} + \mathrm{div}_r \int\limits_{(\infty)} f_2(\vec{r},\vec{v},t)\vec{v}d^3v = 0,$$

$$\frac{\partial f_2(\vec{r},\vec{v},t)}{\partial t} + \mathrm{div}_r\left[f_2(\vec{r},\vec{v},t)\vec{v}\right] + \mathrm{div}_v \int\limits_{(\infty)} f_3(\vec{r},\vec{v},\dot{\vec{v}},t)\dot{\vec{v}}d^3\dot{v} = 0, \qquad (1.1)$$

$$...$$

$$\frac{\partial f_n\left(\vec{r},\vec{v},...,\overset{(n-1)}{\vec{r}},t\right)}{\partial t} + \mathrm{div}_r\left[f_n\left(\vec{r},\vec{v},...,\overset{(n-1)}{\vec{r}},t\right)\vec{v}\right] + ... + \mathrm{div}_{\overset{(n-1)}{r}} \int\limits_{(\infty)} f_{n+1}\left(\vec{r},\vec{v},...,\overset{(n)}{\vec{r}},t\right)\overset{(n)}{\vec{r}}d^3\overset{(n)}{\vec{r}} = 0,$$

$$...$$

The probability density functions satisfy the relations [15,16]

$$f_0(t) \overset{\mathrm{det}}{=} N(t) = \int\limits_{(\infty)} f_1(\vec{r},t)d^3r = \int\limits_{(\infty)}\int\limits_{(\infty)} f_2(\vec{r},\vec{v},t)d^3rd^3v = ... =$$

$$= ... = \int\limits_{(\infty)}\int\limits_{(\infty)}\int\limits_{(\infty)} ... f(\vec{r},\vec{v},\dot{\vec{v}},...,t)d^3rd^3vd^3\dot{v}... \qquad (1.2)$$

where $N(t)$ is normalization factor or a number of particles, which can be non-integer in general case [15,16].

**Definition 1.** *Let us define the average values* $\langle\vec{r}\rangle, \langle\vec{v}\rangle, \langle\langle\vec{v}\rangle\rangle, \langle\dot{\vec{v}}\rangle, \langle\langle\dot{\vec{v}}\rangle\rangle, \langle\langle\langle\dot{\vec{v}}\rangle\rangle\rangle, ...$ *as*

$$N(t)\langle\vec{r}\rangle(t) \overset{\mathrm{det}}{=} \int\limits_{(\infty)} f_1(\vec{r},t)\vec{r}d^3r,$$

$$f_1(\vec{r},t)\langle\vec{v}\rangle(\vec{r},t) \overset{\mathrm{det}}{=} \int\limits_{(\infty)} f_2(\vec{r},\vec{v},t)\vec{v}d^3v,$$

$$N(t)\langle\langle\vec{v}\rangle\rangle(t) \overset{\mathrm{det}}{=} \int\limits_{(\infty)}\int\limits_{(\infty)} f_2(\vec{r},\vec{v},t)\vec{v}d^3rd^3v, , \qquad (1.3)$$

$$f_2(\vec{r},\vec{v},t)\langle\dot{\vec{v}}\rangle(\vec{r},\vec{v},t) \overset{\mathrm{det}}{=} \int\limits_{(\infty)} f_3(\vec{r},\vec{v},\dot{\vec{v}},t)\dot{\vec{v}}d^3\dot{v},$$

$$f_1(r,t)\langle\langle\dot{\vec{v}}\rangle\rangle(\vec{r},t) \overset{\mathrm{det}}{=} \int\limits_{(\infty)}\int\limits_{(\infty)} f_3(\vec{r},\vec{v},\dot{\vec{v}},t)\dot{\vec{v}}d^3vd^3\dot{v},$$

$$N(t)\langle\langle\langle\dot{\vec{v}}\rangle\rangle\rangle(t) \overset{\mathrm{det}}{=} \int\limits_{(\infty)}\int\limits_{(\infty)}\int\limits_{(\infty)} f_3(\vec{r},\vec{v},\dot{\vec{v}},t)\dot{\vec{v}}d^3rd^3vd^3\dot{v},$$

$$...$$

**Definition 2**. *Let us define the functions of entropy* $S_n$, *average entropy* $H_n$ *and source function* $Q_n$ *as*

$$S_n(\vec{r},\vec{v},...,t) \stackrel{det}{=} -\ln f_n(\vec{r},\vec{v},...,t),$$

$$H_1(t) \stackrel{det}{=} -\frac{1}{N}\int_{(\infty)} f_1(\vec{r},t)\ln f_1(\vec{r},t)d^3r = \frac{1}{N}\int_{(\infty)} f_1(\vec{r},t) S_1(\vec{r},t)d^3r = \langle S_1\rangle(t),$$

$$H_2(t) \stackrel{det}{=} -\frac{1}{N}\int_{(\infty)}\int_{(\infty)} f_2(\vec{r},\vec{v},t)\ln f_2(\vec{r},\vec{v},t)d^3rd^3v = \langle\langle S_2\rangle\rangle(t), \qquad (1.4)$$

...

$$H_n(t) \stackrel{det}{=} \langle...\langle S_n\rangle...\rangle(t),$$

...

$$Q_1 \stackrel{det}{=} \mathrm{div}_r\langle\vec{v}\rangle,\ Q_2 \stackrel{det}{=} \mathrm{div}_v\langle\dot{\vec{v}}\rangle,\ ...,\ Q_n \stackrel{det}{=} \mathrm{div}_{r}^{(n-1)}\left\langle \stackrel{(n)}{\vec{r}}\right\rangle.$$

For compact recording of the Vlasov equation chain (1.1), we introduce differential operator $\Pi_n$.

**Definition 3.** *Let us define differential operator* $\Pi_n \stackrel{det}{=} \dfrac{d_n}{dt}$ *of the form:*

$$\Pi_0 = \frac{d_0}{dt} \stackrel{det}{=} \frac{\partial}{\partial t} = \frac{d}{dt},$$

$$\Pi_1 = \frac{d_1}{dt} \stackrel{det}{=} \frac{\partial}{\partial t} + (\langle\vec{v}\rangle,\nabla_r),$$

$$\Pi_2 = \frac{d_2}{dt} \stackrel{det}{=} \frac{\partial}{\partial t} + (\vec{v},\nabla_r) + (\langle\dot{\vec{v}}\rangle,\nabla_v), \qquad (1.5)$$

...

$$\Pi_n = \frac{d_n}{dt} \stackrel{det}{=} \frac{\partial}{\partial t} + (\dot{\vec{r}},\nabla_r) + ... + \left(\stackrel{(n-1)}{\vec{r}},\nabla_{\stackrel{(n-2)}{r}}\right) + \left(\left\langle\stackrel{(n)}{\vec{r}}\right\rangle,\nabla_{\stackrel{(n-1)}{r}}\right) = \frac{\partial}{\partial t} + (\vec{u}_n,\nabla_{\xi_n}),$$

$$\vec{u}_n = \left\{\dot{\vec{r}},\ddot{\vec{r}},...,\left\langle\stackrel{(n)}{\vec{r}}\right\rangle\right\}^T = \left\{\vec{v},\dot{\vec{v}},...,\left\langle\stackrel{(n-1)}{\vec{v}}\right\rangle\right\}^T,\ \nabla_{\xi_n} = \left\{\nabla_r,\nabla_{\dot{r}},...,\nabla_{\stackrel{(n-1)}{r}}\right\}^T = \left\{\nabla_r,\nabla_v,...,\nabla_{\stackrel{(n-2)}{v}}\right\}^T,$$

*where averaging is made over the functions* $f_n(\vec{\xi}_n,t) = f_n(\vec{r},\vec{v},\dot{\vec{v}},...,t)$, *satisfying (1.1) according to (1.3).*

**Lemma 1.** *The chain of the Vlasov equation (1.1) can be presented as follows*

$$\Pi_1 S_1 = Q_1,$$
$$\Pi_2 S_2 = Q_2,$$
$$\ldots \qquad (1.6)$$
$$\Pi_n S_n = Q_n,$$
$$\ldots$$

## 2. Chain of entropy equations

Let us show that the chain of equations for entropy $H_n$ results from the chain of the Vlasov equations (1.6).

**Theorem 1.** *Let functions $S_n = -\ln f_n$ satisfy chain of equations (1.6), then the functions of $H_n$ satisfy the chain of equations:*

$$\frac{d}{dt}\left[N(t)H_1(t)\right] = N(t)\langle Q_1 \rangle(t),$$
$$\frac{d}{dt}\left[N(t)H_2(t)\right] = N(t)\langle\langle Q_2 \rangle\rangle(t),$$
$$\ldots \qquad (2.1)$$
$$\frac{d}{dt}\left[N(t)H_n(t)\right] = N(t)\langle\ldots\langle Q_n \rangle\ldots\rangle(t),$$
$$\ldots$$

**Corollary 1.** *If the number of particles is constant, that is $N(t) = const$, then theorem 1 results in the function of entropy $H_n$ satisfying the equation chain*

$$\frac{dH_1}{dt} = \langle Q_1 \rangle,$$
$$\frac{dH_2}{dt} = \langle\langle Q_2 \rangle\rangle,$$
$$\ldots \qquad (2.2)$$
$$\frac{dH_n}{dt} = \langle\ldots\langle Q_n \rangle\ldots\rangle,$$
$$\ldots$$

Equations (2.2) determine the derivative of the average value of $S_n$, we obtain the expressions for average derivative $\Pi_n$ of $S_n$.

**Theorem 2.** Let $N(t) = const$ and functions $S_n = -\ln f_n$ satisfy the chain of the Vlasov equations (1.6), then

$$\frac{d}{dt}\langle...\langle S_n\rangle...\rangle = \langle...\langle \Pi_n S_n\rangle...\rangle. \qquad (2.22)$$

or

$$\frac{dH_n}{dt} = \langle...\langle \Pi_n S_n\rangle...\rangle.$$

### 3. Relations for average velocities and accelerations of higher order

Let us obtain the relations between the average values of the velocities and the total derivatives of the average velocities [19].

**Theorem 3.** If the distribution functions $f_n$ satisfy the chain of the Vlasov equations (1.1), then the following relations are true:

$$\frac{d}{dt}\left[N(t)\langle \vec{r}\rangle(t)\right] = N(t)\langle\langle\vec{v}\rangle\rangle(t),$$

$$\frac{d^2}{dt^2}\left[N(t)\langle \vec{r}\rangle(t)\right] = N(t)\langle\langle\langle\dot{\vec{v}}\rangle\rangle\rangle(t), \qquad (3.1)$$

$$\frac{d^3}{dt^3}\left[N(t)\langle \vec{r}\rangle(t)\right] = N(t)\langle\langle\langle\langle\ddot{\vec{v}}\rangle\rangle\rangle\rangle(t),$$

....

To prove Theorem 3, the following lemma is necessary [19].

**Lemma 2.** The following formula is true:

$$\int\limits_{(\infty)} \vec{r}\cdot div\left[\vec{A}(\vec{r})\right]d^3r = -\int\limits_{(\infty)} \vec{A}(\vec{r})d^3r, \quad for \ \left\|\vec{A}\right\|_\infty = 0 \qquad (3.2)$$

**Theorem 4.** Let the functions $f_n$ of the probability density distribution satisfy the chain of the Vlasov equations (1.1), then

$$f_0\langle\langle\vec{v}\rangle\rangle = f_0 \frac{d_0}{dt}\langle\vec{r}\rangle + \langle\vec{r}\rangle\frac{d_0}{dt}f_0, \qquad (3.10)$$

$$f_1\langle\langle\dot{\vec{v}}\rangle\rangle = f_1 \frac{d_1}{dt}\langle\vec{v}\rangle + \int\limits_{(\infty)} (\vec{v}-\langle\vec{v}\rangle)(\vec{v}-\langle\vec{v}\rangle, \nabla_r f_2)d^3v,$$

$$f_2\langle\langle\ddot{\vec{v}}\rangle\rangle = f_2 \frac{d_2}{dt}\langle\dot{\vec{v}}\rangle + \int\limits_{(\infty)} (\dot{\vec{v}}-\langle\dot{\vec{v}}\rangle)(\dot{\vec{v}}-\langle\dot{\vec{v}}\rangle, \nabla_v f_3)d^3\dot{v},$$

...

or

$$\langle\langle\vec{v}\rangle\rangle - \frac{d_0}{dt}\langle\vec{r}\rangle = \langle\vec{r}\rangle\frac{d_0}{dt}\ln f_0,$$

$$\langle\langle\dot{\vec{v}}\rangle\rangle - \frac{d_1}{dt}\langle\vec{v}\rangle = \frac{\int_{(\infty)}(\vec{v}-\langle\vec{v}\rangle)(\vec{v}-\langle\vec{v}\rangle,\nabla_r f_2)d^3v}{\int_{(\infty)} f_2 d^3v},$$

$$\langle\langle\ddot{\vec{v}}\rangle\rangle - \frac{d_2}{dt}\langle\dot{\vec{v}}\rangle = \frac{\int_{(\infty)}(\dot{\vec{v}}-\langle\dot{\vec{v}}\rangle)(\dot{\vec{v}}-\langle\dot{\vec{v}}\rangle,\nabla_v f_3)d^3\dot{v}}{\int_{(\infty)} f_3 d^3\dot{v}},$$

...

Let us consider the properties of equations (3.10).

***Theorem 5.*** *Let the functions $f_n$ of the probability density distribution satisfy the chain of the Vlasov equations (1.1), then*

$$\langle\langle\vec{v}\rangle\rangle(t) = \frac{d_0}{dt}\langle\vec{r}\rangle(t) + \langle\vec{r}\rangle(t)\frac{d_0}{dt}\ln f_0(t),$$

$$\langle\langle\langle\dot{\vec{v}}\rangle\rangle\rangle(t) = \left\langle\frac{d_1}{dt}\langle\vec{v}\rangle\right\rangle(t), \qquad \int_{(\infty)}\int_{(\infty)}(\vec{v}-\langle\vec{v}\rangle)(\vec{v}-\langle\vec{v}\rangle,\nabla_r f_2)d^3v d^3r = 0, \qquad (3.22)$$

$$\langle\langle\langle\langle\ddot{\vec{v}}\rangle\rangle\rangle\rangle(t) = \left\langle\left\langle\frac{d_2}{dt}\langle\dot{\vec{v}}\rangle\right\rangle\right\rangle(t), \qquad \int_{(\infty)}\int_{(\infty)}\int_{(\infty)}(\dot{\vec{v}}-\langle\dot{\vec{v}}\rangle)(\dot{\vec{v}}-\langle\dot{\vec{v}}\rangle,\nabla_v f_3)d^3r d^3v d^3\dot{v} = 0,$$

... ...

From Theorem 3, 5 it follows that operator $\Pi_n$ has the following properties.

***Corollary 2.*** *Let the functions $f_n$ of the probability density distribution satisfy the chain of the Vlasov equations (1.1) and $N(t) = const$, then*

$$\Pi_0\langle\vec{r}\rangle = \langle\langle\vec{v}\rangle\rangle, \qquad (3.39)$$

$$\Pi_1\langle\vec{v}\rangle = \langle\langle\dot{\vec{v}}\rangle\rangle,$$

$$\Pi_2\langle\dot{\vec{v}}\rangle = \langle\langle\ddot{\vec{v}}\rangle\rangle,$$

...

$$\Pi_n\left\langle\overset{(n)}{\vec{r}}\right\rangle = \left\langle\left\langle\overset{(n+1)}{\vec{r}}\right\rangle\right\rangle,$$

...

***Corollary 3.*** *Let the functions $f_n$ of the probability density distribution satisfy the chain of the Vlasov equations (1.1) and $N(t) = const$, then*

$$\Pi_{n-1}\left\langle\left\langle\vec{r}^{(n)}\right\rangle\right\rangle=\left\langle\Pi_n\left\langle\vec{r}^{(n)}\right\rangle\right\rangle, \quad \Pi_{n-1}\langle S_n\rangle=\langle\Pi_n S_n\rangle. \tag{3.40}$$

For example,

$$\Pi_0\left\langle\left\langle\left\langle\dot{\vec{v}}\right\rangle\right\rangle\right\rangle=\left\langle\Pi_1\left\langle\left\langle\dot{\vec{v}}\right\rangle\right\rangle\right\rangle=\left\langle\left\langle\Pi_2\left\langle\dot{\vec{v}}\right\rangle\right\rangle\right\rangle=\left\langle\left\langle\left\langle\ddot{\vec{v}}\right\rangle\right\rangle\right\rangle,$$

$$\Pi_0\left\langle\langle S_2\rangle\right\rangle=\left\langle\Pi_1\langle S_2\rangle\right\rangle=\left\langle\langle\Pi_2 S_2\rangle\right\rangle.$$

## 4. Cut-off of the Vlasov equation chain

The Vlasov equation chain (1.1) is an infinite one and, for its practical application, it is necessary to cut it off at some equation. The chain cutoff at the second equation for the probability density function $f_2(\vec{r},\vec{v},t)$ is the most common one.

$$\frac{d_2 S_2}{dt}=Q_2 \text{ or } \Pi_2 f_2=-f_2 Q_2 \text{ or } \Pi_2 S_2=Q_2$$
$$\frac{\partial f_2}{\partial t}+\left(\vec{v},\nabla_r f_2\right)+\left(\langle\dot{\vec{v}}\rangle,\nabla_v f_2\right)=-f_2\operatorname{div}_v\left[\langle\dot{\vec{v}}\rangle\right]. \tag{4.1}$$

Let us find value $\langle\dot{\vec{v}}\rangle$.

**Theorem 6.** *Let the distribution functions $f_n$ satisfy (1.1), then for $\langle\dot{\vec{v}}\rangle$ the following expression is true*

$$\langle\dot{\vec{v}}\rangle=\frac{d_1}{dt}\langle\vec{v}\rangle+(\vec{v}-\langle\vec{v}\rangle)(\vec{v}-\langle\vec{v}\rangle,\nabla_r S_2)+\vec{\eta}, \tag{4.2}$$

*where $\vec{\eta}$ is some vector-function, having this property*

$$\int_{(\infty)} f_2\vec{\eta}d^3 v=f_1\langle\vec{\eta}\rangle=0 \Rightarrow \langle\vec{\eta}\rangle=0. \tag{4.3}$$

In paper [22] it was shown that the following equation of motion is true

$$\frac{d_1}{dt}\langle\vec{v}\rangle(\vec{r},t)=\frac{e}{m}\left(\vec{E}(\vec{r},t)+\left[\langle\vec{v}\rangle(\vec{r},t),\vec{B}(\vec{r},t)\right]\right)=\frac{\vec{F}}{m}. \tag{4.6}$$

Using (4.6) in (4.2), we obtain the expression for $\langle\dot{\vec{v}}\rangle$

$$\langle\dot{\vec{v}}\rangle=\frac{e}{m}\left(\vec{E}(\vec{r},t)+\left[\langle\vec{v}\rangle(\vec{r},t),\vec{B}(\vec{r},t)\right]\right)+(\vec{v}-\langle\vec{v}\rangle)(\vec{v}-\langle\vec{v}\rangle,\nabla_r S_2)+\vec{\eta}. \tag{4.7}$$

From (4.7) it follows, that the right side $Q_2$ of the equation (4.1) is of the form:

$$Q_2 = \mathrm{div}_v \langle \dot{\vec{v}} \rangle = \mathrm{div}_v (\vec{v} - \langle \vec{v} \rangle)(\vec{v} - \langle \vec{v} \rangle, \nabla_r S_2) + \mathrm{div}_v \vec{\eta} = (\delta \vec{v}, \nabla_r S_2) \mathrm{div}_v (\delta \vec{v}) +$$
$$+ (\delta \vec{v}, \nabla_v (\delta \vec{v}, \nabla_r S_2)) + \mathrm{div}_v \vec{\eta} = 3(\delta \vec{v}, \nabla_r S_2) + (\delta \vec{v}, \nabla_v (\delta \vec{v}, \nabla_r S_2)) + \mathrm{div}_v \vec{\eta},$$
$$Q_2 = 3G + (\delta \vec{v}, \nabla_v G) + \mathrm{div}_v \vec{\eta}, \tag{4.8}$$

where $\delta \vec{v} \stackrel{\mathrm{det}}{=} \vec{v} - \langle \vec{v} \rangle$, $G \stackrel{\mathrm{det}}{=} (\delta \vec{v}, \nabla_r S_2)$. As a result, the Vlasov equation (4.1) is of the form (modified Vlasov equation):

$$\frac{\partial f_2}{\partial t} + (\vec{v}, \nabla_r f_2) + \left(\frac{\vec{F}}{m}, \nabla_v f_2\right) = -\Lambda f_2, \tag{4.9}$$

where

$$\Lambda = Q_2 + (\delta \vec{v}, \nabla_r S_2)(\delta \vec{v}, \nabla_v S_2) + (\vec{\eta}, \nabla_v S_2). \tag{4.10}$$

Note that in the general case value $\Lambda \neq 0$ (4.10).

***Corollary 4.*** *It follows from Theorem 6, that* $\langle \dot{\vec{v}} \rangle \neq \dfrac{d}{dt} \langle \vec{v} \rangle$. *Consequently, value* $\langle \dot{\vec{v}} \rangle$ *in equation (4.1), according to (4.2), is not acceleration* $\dfrac{d}{dt} \langle \vec{v} \rangle = \dfrac{\vec{F}}{m}$.

***Theorem 7.*** *If the following condition is fulfilled*

$$(\vec{v} - \langle \vec{v} \rangle)(\vec{v} - \langle \vec{v} \rangle, \nabla_r S_2) + \vec{\eta} = 0, \tag{4.11}$$

*then the Vlasov equation (4.1) is of the form*

$$\frac{\partial f_2}{\partial t} + (\vec{v}, \nabla_r f_2) + \left(\frac{\vec{F}}{m}, \nabla_v f_2\right) = 0, \tag{4.12}$$

*which at* $N = const$ *(constant particle number) corresponds to constant entropy* $H_2 = const$, *that is*

$$H_2 = -\int\limits_{(\infty)} \int\limits_{(\infty)} f_2 \ln f_2 d^3 r d^3 v = const. \tag{4.13}$$

**Conclusion**

As it is known from the principle of entropy increase, entropy $H_2$ increases or is constant. From the proved theorems, it follows that substitution of the modified Vlasov equation (4.9)

$$\frac{\partial f_2}{\partial t} + (\vec{v}, \nabla_r f_2) + \left(\frac{\vec{F}}{m}, \nabla_v f_2\right) = -\Lambda f_2,$$

with equation (4.12)

$$\frac{\partial f_2}{\partial t}+(\vec{v},\nabla_r f_2)+\left(\frac{\vec{F}}{m},\nabla_v f_2\right)=0,$$

leads to consideration of the special case only, for which the entropy is constant $H_2 = const$. Thus, only the special case of the modified Vlasov equation (4.9) is considered corresponding to the classical Liouville's equation (4.12). However, as it is seen from (4.9), the modified Vlasov equation differs from the Liouville's equation (4.12) in the general case.

As a result, a significant class of the solutions for equation (4.9), for which entropy $H_2$ increases, is left out of consideration. Considering these particular solutions ($\frac{dH_2}{dt} \geq 0$) is of great importance for the problems of plasma physics, investigation of the stability aspect in the problems of plasma confinement, as well as for controlled thermonuclear fusion initiation.

**Appendices**

*Proof of Lemma 1*

According to Definition 3 and formulae (1.3), equation chain (1.1) is written as follows

$$\frac{\partial f_1}{\partial t}+\operatorname{div}_r\left[f_1\langle\vec{v}\rangle\right]=\frac{\partial f_1}{\partial t}+\left(\langle\vec{v}\rangle,\nabla_r f_1\right)+f_1\operatorname{div}_r\langle\vec{v}\rangle=0,$$

$$\frac{\partial f_2}{\partial t}+\operatorname{div}_r\left[f_2\vec{v}\right]+\operatorname{div}_v\left[f_2\langle\dot{\vec{v}}\rangle\right]=\frac{\partial f_2}{\partial t}+(\vec{v},\nabla_r f_2)+\left(\langle\dot{\vec{v}}\rangle,\nabla_v f_2\right)+f_2\operatorname{div}_v\langle\dot{\vec{v}}\rangle=0, \quad (1.7)$$

...

$$\frac{\partial f_n}{\partial t}+\operatorname{div}_r\left[f_n\vec{v}\right]+...+\operatorname{div}_{r^{(n-1)}}\left[f_n\left\langle\stackrel{(n)}{\vec{r}}\right\rangle\right]=$$

$$=\frac{\partial f_n}{\partial t}+(\vec{v},\nabla_r f_n)+(\dot{\vec{v}},\nabla_v f_n)+...+\left(\left\langle\stackrel{(n)}{\vec{r}}\right\rangle,\nabla_{r^{(n-1)}} f_n\right)+f_n\operatorname{div}_{r^{(n-1)}}\left\langle\stackrel{(n)}{\vec{r}}\right\rangle=0,$$

...

Dividing equations (1.7) by the corresponding functions $f_n$ and taking into account Definitions 2, 3, we obtain expression (1.6), which was to be proved.

*Proof of Theorem 1*

Let us multiply the first equation in (1.1) by $(1+\ln f_1)$ and integrate it over $d^3 r$, we obtain

$$(1+\ln f_1)\frac{\partial f_1}{\partial t}+(1+\ln f_1)\operatorname{div}_r\left[\langle\vec{v}\rangle f_1\right]=0,$$

$$\int_{(\infty)}\frac{\partial}{\partial t}(f_1 \ln f_1)d^3 r+\int_{(\infty)}(1+\ln f_1)\operatorname{div}_r\left[\langle\vec{v}\rangle f_1\right]d^3 r=0. \quad (2.3)$$

The first integral in (2.3) is of the form:

$$\int\limits_{(\infty)} (1+\ln f_1)\frac{\partial f_1}{\partial t} d^3r = \int\limits_{(\infty)} \frac{\partial}{\partial t}(f_1 \ln f_1) d^3r = -\frac{\partial}{\partial t}\int\limits_{(\infty)} f_1 S_1 d^3r = -\frac{d}{dt}\Big[N(t)\langle S_1\rangle(t)\Big]. \qquad (2.4)$$

The second integral in (2.3)

$$\int\limits_{(\infty)} (1+\ln f_1)\operatorname{div}_r\big[\langle\vec{v}\rangle f_1\big]d^3r = \int\limits_{(\infty)} (\langle\vec{v}\rangle,(1+\ln f_1)\nabla_r f_1)d^3r + \int\limits_{(\infty)} (1+\ln f_1)f_1 \operatorname{div}_r\langle\vec{v}\rangle d^3r =$$

$$= \int\limits_{(\infty)} (\langle\vec{v}\rangle,\nabla_r(f_1 \ln f_1))d^3r + \int\limits_{(\infty)} f_1 \operatorname{div}_r\langle\vec{v}\rangle d^3r + \int\limits_{(\infty)} f_1 \ln f_1 \operatorname{div}_r\langle\vec{v}\rangle d^3r =$$

$$= -\int\limits_{(\infty)} (\langle\vec{v}\rangle,\nabla_r(f_1 S_1))d^3r - \int\limits_{(\infty)} f_1 S_1 \operatorname{div}_r\langle\vec{v}\rangle d^3r + \int\limits_{(\infty)} f_1 \operatorname{div}_r\langle\vec{v}\rangle d^3r =$$

$$= -\int\limits_{(\infty)} \operatorname{div}_r\big[f_1 S_1 \langle\vec{v}\rangle\big]d^3r + \int\limits_{(\infty)} f_1 \operatorname{div}_r\langle\vec{v}\rangle d^3r = \int\limits_{(\infty)} f_1 Q_1 d^3r = N(t)\langle Q_1\rangle(t), \qquad (2.5)$$

where the following condition is supposed to be fulfilled
$\int\limits_{(\infty)} \operatorname{div}_r\big[f_1 S_1 \langle\vec{v}\rangle\big]d^3r = \int\limits_{\Sigma_\infty^r} f_1 S_1 \langle\vec{v}\rangle d\vec{\sigma}_r = 0$ (this condition is imposed on the functions $f_n$ as a default at deducing the chain of the Vlasov equations (1.1) [15,16]).

Using (2.4) and (2.5) in (2.3), we obtain ultimately

$$\frac{d}{dt}\Big[N(t)\langle S_1\rangle(t)\Big] = N(t)\langle Q_1\rangle(t). \qquad (2.6)$$

If the number of particles is constant, then $N(t) = N_0 = const$, and expression (2.6) is as follows

$$\frac{d\langle S_1\rangle}{dt} = \langle Q_1\rangle. \qquad (2.7)$$

By virtue of Definition 2 (1.2), equation (2.6) coincides with the first equation from (2.1), and equation (2.7) coincides with the first equation from (2.2).
Let us do analogue computation for the second equation in chain (1.1), we obtain

$$(1+\ln f_2)\frac{\partial f_2}{\partial t} + (1+\ln f_2)\operatorname{div}_r\big[\vec{v}f_2\big] + (1+\ln f_2)\operatorname{div}_v\Big[\langle\dot{\vec{v}}\rangle f_2\Big] = 0,$$

$$\int\limits_{(\infty)}\int\limits_{(\infty)} \frac{\partial}{\partial t}(f_2 \ln f_2) d^3r d^3v + \int\limits_{(\infty)}\int\limits_{(\infty)} (1+\ln f_2)\operatorname{div}_r\big[\vec{v}f_2\big] d^3r d^3v + \qquad (2.8)$$

$$+\int\limits_{(\infty)}\int\limits_{(\infty)} (1+\ln f_2)\operatorname{div}_v\Big[\langle\dot{\vec{v}}\rangle f_2\Big] d^3r d^3v = 0.$$

The first integral in (2.8) is of the form:

$$\int\limits_{(\infty)}\int\limits_{(\infty)} (1+\ln f_2)\frac{\partial f_2}{\partial t} d^3r d^3v = \int\limits_{(\infty)}\int\limits_{(\infty)} \frac{\partial}{\partial t}(f_2 \ln f_2) d^3r d^3v = -\frac{\partial}{\partial t}\int\limits_{(\infty)}\int\limits_{(\infty)} f_2 S_2 d^3r d^3v =$$

$$= -\frac{d}{dt}\Big[N(t)\langle\langle S_2\rangle\rangle(t)\Big]. \qquad (2.9)$$

The second integral in (2.8)

$$\int_{(\infty)}\int_{(\infty)} (1+\ln f_2)\operatorname{div}_r[\vec{v}f_2]d^3rd^3v = \int_{(\infty)}\int_{(\infty)} (\vec{v},(1+\ln f_2)\nabla_r f_2)d^3rd^3v +$$
$$+ \int_{(\infty)}\int_{(\infty)} (1+\ln f_2)f_2 \operatorname{div}_r \vec{v}d^3rd^3v = \int_{(\infty)}\int_{(\infty)} (\vec{v},\nabla_r(f_2 \ln f_2))d^3rd^3v =$$
$$= -\int_{(\infty)}\int_{(\infty)} (\vec{v},\nabla_r(f_2 S_2))d^3rd^3v = -\int_{(\infty)}\int_{(\infty)} \operatorname{div}_r[f_2 S_2 \vec{v}]d^3rd^3v =$$
$$= -\int_{(\infty)}\int_{\Sigma^r_\infty} f_2 S_2 \vec{v}d\vec{\sigma}_r d^3v = 0,$$

(2.10)

where the condition $\int_{\Sigma^r_\infty} f_2 S_2 \vec{v}d\vec{\sigma}_r = 0$ is supposed to be fulfilled. The third integral in (2.8)

$$\int_{(\infty)}\int_{(\infty)} (1+\ln f_2)\operatorname{div}_v\left[\langle\dot{\vec{v}}\rangle f_2\right]d^3rd^3v = \int_{(\infty)}\int_{(\infty)} (1+\ln f_2)\left[f_2 \operatorname{div}_v\langle\dot{\vec{v}}\rangle + (\langle\dot{\vec{v}}\rangle,\nabla_v f_2)\right]d^3rd^3v =$$
$$= \int_{(\infty)}\int_{(\infty)} \left[f_2(1+\ln f_2)\operatorname{div}_v\langle\dot{\vec{v}}\rangle + (\langle\dot{\vec{v}}\rangle,(1+\ln f_2)\nabla_v f_2)\right]d^3rd^3v =$$
$$= \int_{(\infty)}\int_{(\infty)} \left[f_2 \operatorname{div}_v\langle\dot{\vec{v}}\rangle + f_2 \ln f_2 \operatorname{div}_v\langle\dot{\vec{v}}\rangle + (\langle\dot{\vec{v}}\rangle,\nabla_v f_2 \ln f_2)\right]d^3rd^3v =$$
$$= \int_{(\infty)}\int_{(\infty)} f_2 \operatorname{div}_v\langle\dot{\vec{v}}\rangle d^3rd^3v - \int_{(\infty)}\int_{(\infty)} \left[(\langle\dot{\vec{v}}\rangle,\nabla_v f_2 S_2) + f_2 S_2 \operatorname{div}_v\langle\dot{\vec{v}}\rangle\right]d^3rd^3v =$$
$$= N(t)\langle\langle Q_2\rangle\rangle(t) - \int_{(\infty)}\int_{(\infty)} \operatorname{div}_v\left[\langle\dot{\vec{v}}\rangle f_2 S_2\right]d^3rd^3v = N(t)\langle\langle Q_2\rangle\rangle(t) - \int_{(\infty)}\int_{\Sigma^v_\infty} \langle\dot{\vec{v}}\rangle f_2 S_2 d^3\vec{\sigma}_v d^3r =$$
$$= N(t)\langle\langle Q_2\rangle\rangle(t),$$

(2.11)

where the condition $\int_{\Sigma^v_\infty} \langle\dot{\vec{v}}\rangle f_2 S_2 d^3\vec{\sigma}_v = 0$ is supposed to be fulfilled. Using (2.9)-(2.11) in (2.8), we obtain ultimately

$$\frac{d}{dt}\left[N(t)\langle\langle S_2\rangle\rangle(t)\right] = N(t)\langle\langle Q_2\rangle\rangle(t). \qquad (2.12)$$

For $S_3, S_4,...$ one may obtain the expressions analogue to expressions (2.6) and (2.12). In the general case for the equation

$$\frac{\partial f_n}{\partial t} + \operatorname{div}_{\xi_n}[\vec{u}_n f_n] = 0, \qquad (2.13)$$

we obtain

$$(1+\ln f_n)\frac{\partial f_n}{\partial t} + (1+\ln f_n)\operatorname{div}_{\xi_n}[\vec{u}_n f_n] = 0,$$
$$\int_{(\infty)}...\int_{(\infty)} \frac{\partial}{\partial t}(f_n \ln f_n)d^3r...d^{3(n-1)}r + \int_{(\infty)}...\int_{(\infty)} (1+\ln f_n)\operatorname{div}_{\xi_n}[\vec{u}_n f_n]d^3r...d^{3(n-1)}r = 0,$$

(2.14)

The first integral in (2.14) is of the form:

$$\int_{(\infty)}...\int_{(\infty)} \frac{\partial}{\partial t}(f_n \ln f_n) d^3r...d^{3}\overset{(n-1)}{r} = -\frac{\partial}{\partial t}\int_{(\infty)}...\int_{(\infty)} f_n S_n d^3r...d^{3}\overset{(n-1)}{r} = -\frac{d}{dt}\left[N(t)\langle...\langle S_n\rangle...\rangle(t)\right].$$
(2.15)

Let us rearrange the expression $\text{div}_{\xi_n}[\vec{u}_n f_n]$. Taking into account that

$$\text{div}_r[\vec{v}] = \text{div}_v[\dot{\vec{v}}] = ... = \text{div}_{\underset{\vec{v}}{(n-2)}}\left[\left\langle \overset{(n-1)}{\vec{v}} \right\rangle\right] = 0,$$
(2.16)

we obtain

$$\text{div}_{\xi_n}[\vec{u}_n f_n] = \text{div}_r[\vec{v} f_n] + \text{div}_v[\dot{\vec{v}} f_n] + ... + \text{div}_{\underset{\vec{v}}{(n-2)}}\left[\left\langle \overset{(n-1)}{\vec{v}} \right\rangle f_n\right] = (\vec{v}, \nabla_r f_n) + (\dot{\vec{v}}, \nabla_v f_n) + ... +$$

$$+ \left(\left\langle \overset{(n-1)}{\vec{v}} \right\rangle, \nabla_{\underset{\vec{v}}{(n-2)}} f_n\right) + f_n \text{div}_{\underset{\vec{v}}{(n-2)}}\left[\left\langle \overset{(n-1)}{\vec{v}} \right\rangle\right],$$
(2.17)

multiplying (2.17) by $(1 + \ln f_n)$ and taking into account (2.16), we obtain

$$(1 + \ln f_n)\text{div}_{\xi_n}[\vec{u}_n f_n] = (\vec{v},(1 + \ln f_n)\nabla_r f_n) + (\dot{\vec{v}},(1 + \ln f_n)\nabla_v f_n) + ... +$$

$$+ \left(\left\langle \overset{(n-1)}{\vec{v}} \right\rangle,(1 + \ln f_n)\nabla_{\underset{\vec{v}}{(n-2)}} f_n\right) + (1 + \ln f_n) f_n \text{div}_{\underset{\vec{v}}{(n-2)}}\left[\left\langle \overset{(n-1)}{\vec{v}} \right\rangle\right] = (\vec{v}, \nabla_r f_n \ln f_n) +$$

$$+ (\dot{\vec{v}}, \nabla_v f_n \ln f_n) + ... + \left(\left\langle \overset{(n-1)}{\vec{v}} \right\rangle, \nabla_{\underset{\vec{v}}{(n-2)}} f_n \ln f_n\right) + f_n \text{div}_{\underset{\vec{v}}{(n-2)}}\left[\left\langle \overset{(n-1)}{\vec{v}} \right\rangle\right] +$$

$$+ f_n \ln f_n \text{div}_{\underset{\vec{v}}{(n-2)}}\left[\left\langle \overset{(n-1)}{\vec{v}} \right\rangle\right] = -(\vec{v}, \nabla_r f_n S_n) - (\dot{\vec{v}}, \nabla_v f_n S_n) - ... - \left(\left\langle \overset{(n-1)}{\vec{v}} \right\rangle, \nabla_{\underset{\vec{v}}{(n-2)}} f_n S_n\right) +$$

$$+ f_n \text{div}_{\underset{\vec{v}}{(n-2)}}\left[\left\langle \overset{(n-1)}{\vec{v}} \right\rangle\right] - f_n S_n \text{div}_{\underset{\vec{v}}{(n-2)}}\left[\left\langle \overset{(n-1)}{\vec{v}} \right\rangle\right] = -\text{div}_r[\vec{v} f_n S_n] - \text{div}_v[\dot{\vec{v}} f_n S_n] - ... -$$

$$- \text{div}_{\underset{\vec{v}}{(n-2)}}\left[\left\langle \overset{(n-1)}{\vec{v}} \right\rangle f_n S_n\right] + f_n \text{div}_{\underset{\vec{v}}{(n-2)}}\left[\left\langle \overset{(n-1)}{\vec{v}} \right\rangle\right].$$
(2.18)

Using (2.18) in the second integral of equation (2.14)

$$\int\limits_{(\infty)}...\int\limits_{(\infty)} (1+\ln f_n) \mathrm{div}_{\xi_n} [\vec{u}_n f_n] d^3r...d^3\overset{(n-1)}{r} = -\int\limits_{(\infty)}...\int\limits_{(\infty)} \mathrm{div}_r [\vec{v} f_n S_n] d^3r...d^3\overset{(n-1)}{r} -$$

$$-\int\limits_{(\infty)}...\int\limits_{(\infty)} \mathrm{div}_v [\dot{\vec{v}} f_n S_n] d^3r...d^3\overset{(n-1)}{r} -...-\int\limits_{(\infty)}...\int\limits_{(\infty)} \mathrm{div}_{\underset{\vec{v}}{(n-2)}} \left[\left\langle \overset{(n-1)}{\vec{v}} \right\rangle f_n S_n \right] d^3r...d^3\overset{(n-1)}{r} + \quad (2.19)$$

$$+\int\limits_{(\infty)}...\int\limits_{(\infty)} f_n \mathrm{div}_{\underset{\vec{v}}{(n-2)}} \left[\left\langle \overset{(n-1)}{\vec{v}} \right\rangle \right] d^3r...d^3\overset{(n-1)}{r} = \int\limits_{(\infty)}...\int\limits_{(\infty)} f_n Q_n d^3r...d^3\overset{(n-1)}{r} = N(t)\langle...\langle Q_n\rangle...\rangle(t).$$

Using (2.15) and (2.19) in (2.14), we obtain ultimately

$$\frac{d}{dt}\big[N(t)\langle...\langle S_n\rangle...\rangle(t)\big] = N(t)\langle...\langle Q_n\rangle...\rangle(t). \quad (2.20)$$

If the number of particles is constant, then $N(t) = N_0 = const$, expression (2.20) is of the form

$$\frac{d}{dt}\langle...\langle S_n\rangle...\rangle(t) = \langle...\langle Q_n\rangle...\rangle(t), \quad (2.21)$$

which was to be proved.

*Proof of Theorem 2*

Let us multiply the equation for $S_n$ in (1.6) by $f_n$ and integrate it, we obtain

$$\int\limits_{(\infty)}...\int\limits_{(\infty)} \frac{d_n S_n}{dt} f_n d^3r...d^3\overset{(n-1)}{r} = \int\limits_{(\infty)}...\int\limits_{(\infty)} f_n Q_n d^3r...d^3\overset{(n-1)}{r},$$

$$\left\langle...\left\langle \frac{d_n S_n}{dt}\right\rangle...\right\rangle(t) = \langle...\langle Q_n\rangle...\rangle(t). \quad (2.23)$$

Comparing (2.23) and (2.2) shows that at $N(t) = const$ the derivative of the average equals to the average of the derivative of $\Pi_n$ (2.22), which was to be proved.

*Proof of Lemma 2.*

We write expression (3.2) in the Cartesian coordinate system:

$$\int\limits_{(\infty)} \vec{r} \cdot \mathrm{div}\big[\vec{A}(\vec{r})\big] d^3r =$$

$$= \vec{e}_x \int\limits_{-\infty}^{+\infty}\int\limits_{-\infty}^{+\infty}\int\limits_{-\infty}^{+\infty} x\left(\frac{\partial A^{(x)}(x,y,z)}{\partial x} + \frac{\partial A^{(y)}(x,y,z)}{\partial y} + \frac{\partial A^{(z)}(x,y,z)}{\partial z}\right) dxdydz + \quad (3.3)$$

$$+\vec{e}_y \int_{-\infty}^{+\infty}\int_{-\infty}^{+\infty}\int_{-\infty}^{+\infty} y\left(\frac{\partial A^{(x)}(x,y,z)}{\partial x}+\frac{\partial A^{(y)}(x,y,z)}{\partial y}+\frac{\partial A^{(z)}(x,y,z)}{\partial z}\right)dxdydz+$$

$$+\vec{e}_z \int_{-\infty}^{+\infty}\int_{-\infty}^{+\infty}\int_{-\infty}^{+\infty} z\left(\frac{\partial A^{(x)}(x,y,z)}{\partial x}+\frac{\partial A^{(y)}(x,y,z)}{\partial y}+\frac{\partial A^{(z)}(x,y,z)}{\partial z}\right)dxdydz.$$

Let us write each integral in (3.3) separately, starting with the first one:

$$\int_{-\infty}^{+\infty}\int_{-\infty}^{+\infty}\int_{-\infty}^{+\infty} x\left(\frac{\partial A^{(x)}(x,y,z)}{\partial x}+\frac{\partial A^{(y)}(x,y,z)}{\partial y}+\frac{\partial A^{(z)}(x,y,z)}{\partial z}\right)dxdydz =$$

$$= \int_{-\infty}^{+\infty} dz \int_{-\infty}^{+\infty} dy \int_{-\infty}^{+\infty} x\frac{\partial A^{(x)}(x,y,z)}{\partial x}dx + \int_{-\infty}^{+\infty} xdx \int_{-\infty}^{+\infty} dz \int_{-\infty}^{+\infty} \frac{\partial A^{(y)}(x,y,z)}{\partial y}dy + \quad (3.4)$$

$$+\int_{-\infty}^{+\infty} xdx \int_{-\infty}^{+\infty} dy \int_{-\infty}^{+\infty} \frac{\partial A^{(z)}(x,y,z)}{\partial z}dz =$$

$$= \int_{-\infty}^{+\infty} dz \int_{-\infty}^{+\infty} dy \left( xA^{(x)}(x,y,z)\Big|_{x=-\infty}^{x=+\infty} - \int_{-\infty}^{+\infty} A^{(x)}(x,y,z)dx \right) +$$

$$+\int_{-\infty}^{+\infty} xdx \int_{-\infty}^{+\infty} dz \left( A^{(y)}(x,y,z)\Big|_{y=-\infty}^{y=+\infty} \right) +$$

$$+\int_{-\infty}^{+\infty} xdx \int_{-\infty}^{+\infty} dy \left( A^{(z)}(x,y,z)\Big|_{z=-\infty}^{z=+\infty} \right) = -\int_{-\infty}^{+\infty}\int_{-\infty}^{+\infty}\int_{-\infty}^{+\infty} A^{(x)}(x,y,z)dxdydz.$$

For the second and third integrals in (3.3) analogue to (3.4), we obtain:

$$\int_{(\infty)} y\, div\left[\vec{A}(\vec{r})\right]d^3r = -\int_{(\infty)} A^{(y)}(x,y,z)d^3r,$$

$$\int_{(\infty)} z\, div\left[\vec{A}(\vec{r})\right]d^3r = -\int_{(\infty)} A^{(z)}(x,y,z)d^3r. \quad (3.5)$$

Using (3.4)–(3.5) in (3.3), we obtain (3.2), which was to be proved.

***Proof of Theorem 3.***

Testing the first expression in (3.1) and taking into account the first equation in the Vlasov chain and formula (3.2), we obtain:

$$N(t)\langle\vec{r}\rangle(t) = \int_{(\infty)} \vec{r}f_1(\vec{r},t)d^3r,$$

$$\frac{d}{dt}\left[N(t)\langle\vec{r}\rangle(t)\right] = \int_{(\infty)} \vec{r}\frac{\partial f_1(\vec{r},t)}{\partial t}d^3r = -\int_{(\infty)} \vec{r}\,div_r\left[\int_{(\infty)} \vec{v}f_2(\vec{r},\vec{v},t)d^3v\right]d^3r = \quad (3.6)$$

$$= -\int_{(\infty)} \vec{r}\,div_r\left[\langle\vec{v}\rangle(\vec{r},t)f_1(\vec{r},t)\right]d^3r = \int_{(\infty)} \langle\vec{v}\rangle(\vec{r},t)f_1(\vec{r},t)d^3r = N(t)\langle\langle\vec{v}\rangle\rangle(t),$$

The obtained result proves the first identical relation in (3.1). Let us test the second identical relation in (3.1). In order to do it, note that:

$$\frac{d^2}{dt^2}\big[N(t)\langle\vec{r}\rangle(t)\big] = \frac{d}{dt}\left(\frac{d}{dt}\big[N(t)\langle\vec{r}\rangle(t)\big]\right) = \frac{d}{dt}\big[N(t)\langle\langle\vec{v}\rangle\rangle(t)\big]. \qquad (3.7)$$

Consequently, it suffices to show that

$$\frac{d}{dt}\big[N(t)\langle\langle\vec{v}\rangle\rangle(t)\big] = N(t)\langle\langle\langle\dot{\vec{v}}\rangle\rangle\rangle(t). \qquad (3.8)$$

Taking account for the second equation in the Vlasov chain for the function $f_2(\vec{r},\vec{v},t)$ and formula (3.2), we obtain for expression (3.8):

$$N(t)\langle\langle\vec{v}\rangle\rangle(t) = \int_{(\infty)}\int_{(\infty)} \vec{v} f_2(\vec{r},\vec{v},t) d^3v d^3r,$$

$$\frac{d}{dt}\big[N(t)\langle\langle\vec{v}\rangle\rangle(t)\big] = \int_{(\infty)}\int_{(\infty)} \vec{v}\,\frac{\partial f_2(\vec{r},\vec{v},t)}{\partial t} d^3v d^3r =$$

$$= -\int_{(\infty)}\int_{(\infty)} \vec{v}\left\{\mathrm{div}_r\big[f_2(\vec{r},\vec{v},t)\vec{v}\big] + \mathrm{div}_v \int_{(\infty)} \dot{\vec{v}} f_3(\vec{r},\vec{v},\dot{\vec{v}},t) d^3\dot{v}\right\} d^3v d^3r =$$

$$= -\int_{(\infty)} \mathrm{div}_r\big[f_2(\vec{r},\vec{v},t)\vec{v}\big] d^3r \int_{(\infty)} \vec{v} d^3v - \int_{(\infty)}\int_{(\infty)} \vec{v}\,\mathrm{div}_v\big[f_2(\vec{r},\vec{v},t)\langle\dot{\vec{v}}\rangle(\vec{r},\vec{v},t)\big] d^3v d^3r =$$

$$= -\int_{\Sigma_\infty} f_2(\vec{r},\vec{v},t) dS_r \int_{(\infty)} \vec{v} d^3v - \int_{(\infty)} d^3r \int_{(\infty)} \vec{v}\,\mathrm{div}_v\big[f_2(\vec{r},\vec{v},t)\langle\dot{\vec{v}}\rangle(\vec{r},\vec{v},t)\big] d^3v =$$

$$= \int_{(\infty)} d^3r \int_{(\infty)} f_2(\vec{r},\vec{v},t)\langle\dot{\vec{v}}\rangle(\vec{r},\vec{v},t) d^3v = N(t)\langle\langle\langle\dot{\vec{v}}\rangle\rangle\rangle(t). \qquad (3.9)$$

From (3.7)–(3.9) it follows that the second identical relation in expression (3.1) is true. The rest expressions are proved in the same manner with the use of the following equations from the chain of the Vlasov equations. Thus, Theorem 3 is proved.

*Proof of Theorem 4*
Let us test the first expression in (3.10)

$$\frac{d}{dt}\big[N(t)\langle\vec{r}\rangle(t)\big] = \frac{d}{dt}\int_{(\infty)} \vec{r} f_1(\vec{r},t) d^3r. \qquad (3.11)$$

Note that the integral $\int_{(\infty)} \vec{r} f_1(\vec{r},t) d^3r$ is the function of the variable – «$t$», therefore, taking its total derivative $\frac{d}{dt}$ under the integral, one may transit to the partial time derivative $\frac{\partial}{\partial t}$, that is

$$\frac{d}{dt}\int_{(\infty)} \vec{r} f_1(\vec{r},t) d^3r = \int_{(\infty)} \vec{r}\,\frac{\partial}{\partial t} f_1(\vec{r},t) d^3r. \qquad (3.12)$$

Note that a transition of the kind of (3.12) is used, for instance, at deducing the continuity equation, when the change of mass $\frac{d}{dt}M(t)$ is written by integral with respect to the volume of the density $\frac{d}{dt}\int \rho(\vec{r},t)d^3r = \int \frac{\partial \rho}{\partial t}d^3r$. That is why it is $\frac{\partial f_1}{\partial t}$ what is under the integral in expression (3.12) but not $\frac{\partial f_1}{\partial t} + (\langle \vec{v} \rangle, \nabla_r)$, as one may suppose relying on the definition of the total derivative (1.5). According to the above mentioned and taking into account (3.12), (3.2) and the first Vlasov equation (1.1):

$$\frac{\partial f_1}{\partial t} + \mathrm{div}_r\left[\langle \vec{v} \rangle f_1\right] = 0, \qquad (3.13)$$

expression (3.11) is of the following form

$$\frac{d}{dt}\left[N(t)\langle \vec{r} \rangle(t)\right] = -\int_{(\infty)} \vec{r}\,\mathrm{div}_r\left[f_1\langle \vec{v} \rangle\right]d^3r = \int_{(\infty)} f_1\langle \vec{v} \rangle d^3r = N(t)\langle\langle \vec{v} \rangle\rangle(t), \qquad (3.14)$$

Thus, considering (3.14), expression (3.11) is of the form:

$$N(t)\langle\langle \vec{v} \rangle\rangle(t) = N(t)\frac{d}{dt}\langle \vec{r} \rangle(t) + \langle \vec{r} \rangle(t)\frac{dN(t)}{dt},$$

where $f_0(t) \equiv N(t)$, which corresponds to the first relation in (3.1) and proves the first expression in (3.10) is true.

Let us show that the second expression in (3.10) is true. On the one hand,

$$\frac{d_1}{dt}\left[f_1(\vec{r},t)\langle \vec{v} \rangle(\vec{r},t)\right] = \langle \vec{v} \rangle(\vec{r},t)\frac{d_1 f_1(\vec{r},t)}{dt} + f_1(\vec{r},t)\frac{d_1}{dt}\langle \vec{v} \rangle(\vec{r},t). \qquad (3.15)$$

on the other hand,

$$\frac{d_1}{dt}\left[f_1(\vec{r},t)\langle \vec{v} \rangle(\vec{r},t)\right] = \frac{d_1}{dt}\int f_2 \vec{v} d^3v = \int \vec{v}\frac{d_1}{dt}f_2 d^3v. \qquad (3.16)$$

Note that the derivative $\frac{d_1}{dt}$ of the function $f_2(\vec{r},\vec{v},t)$ in expression (3.16) equals to $\frac{d_1}{dt}f_2 = \frac{\partial f_2}{\partial t} + (\langle \vec{v} \rangle, \nabla_r f_2)$, as integrating is taken over variable $\vec{v}$, therefore differencing over $\vec{v}$ is not performed. Equating (3.15) to (3.16), we obtain

$$f_1\frac{d_1}{dt}\langle \vec{v} \rangle = \int \vec{v}\frac{d_1}{dt}f_2 d^3v - \langle \vec{v} \rangle\frac{d}{dt}f_1 = \int \vec{v}\frac{d_1}{dt}f_2 d^3v - \langle \vec{v} \rangle\frac{d}{dt}\int f_2 d^3v = \int \vec{v}\frac{d_1}{dt}f_2 d^3v -$$
$$-\int \langle \vec{v} \rangle \frac{d_1}{dt}f_2 d^3v = \int (\vec{v} - \langle \vec{v} \rangle)\frac{d_1}{dt}f_2 d^3v = \int (\vec{v} - \langle \vec{v} \rangle)\frac{\partial f_2}{\partial t}d^3v + \int (\vec{v} - \langle \vec{v} \rangle)(\langle \vec{v} \rangle, \nabla_r f_2)d^3v. \qquad (3.17)$$

With the use of the second Vlasov equation (1.1)

$$\frac{\partial f_2}{\partial t} + \mathrm{div}_r\left[f_2\vec{v}\right] + \mathrm{div}_v\left[\langle\dot{\vec{v}}\rangle f_2\right] = 0, \qquad (3.18)$$

expression (3.17) is of the form

$$f_1\frac{d_1}{dt}\langle\vec{v}\rangle = -\int(\vec{v}-\langle\vec{v}\rangle)\left(\mathrm{div}_r\left[f_2\vec{v}\right] + \mathrm{div}_v\left[\langle\dot{\vec{v}}\rangle f_2\right]\right)d^3v + \int(\vec{v}-\langle\vec{v}\rangle)(\langle\vec{v}\rangle,\nabla_r f_2)d^3v =$$

$$= -\int(\vec{v}-\langle\vec{v}\rangle)(\vec{v},\nabla_r f_2)d^3v - \int(\vec{v}-\langle\vec{v}\rangle)\mathrm{div}_v\left[\langle\dot{\vec{v}}\rangle f_2\right]d^3v + \int(\vec{v}-\langle\vec{v}\rangle)(\langle\vec{v}\rangle,\nabla_r f_2)d^3v =$$

$$= \int(\vec{v}-\langle\vec{v}\rangle)(\langle\vec{v}\rangle-\vec{v},\nabla_r f_2)d^3v - \int\vec{v}\,\mathrm{div}_v\left[\langle\dot{\vec{v}}\rangle f_2\right]d^3v + \langle\vec{v}\rangle\int\mathrm{div}_v\left[\langle\dot{\vec{v}}\rangle f_2\right]d^3v =$$

$$= \int(\vec{v}-\langle\vec{v}\rangle)(\langle\vec{v}\rangle-\vec{v},\nabla_r f_2)d^3v + \int\langle\dot{\vec{v}}\rangle f_2 d^3v =$$

$$= f_1\langle\langle\dot{\vec{v}}\rangle\rangle - \int(\vec{v}-\langle\vec{v}\rangle)(\vec{v}-\langle\vec{v}\rangle,\nabla_r f_2)d^3v,$$

that is

$$f_1\langle\langle\dot{\vec{v}}\rangle\rangle = f_1\frac{d_1}{dt}\langle\vec{v}\rangle + \int(\vec{v}-\langle\vec{v}\rangle)(\vec{v}-\langle\vec{v}\rangle,\nabla_r)f_2 d^3v,$$

which proves the second expression in (3.10) is true.

Let us check the third expression in (3.10). Doing computations analogue to the above described, we obtain

$$\frac{d_2}{dt}\left[f_2\langle\dot{\vec{v}}\rangle\right] = \langle\dot{\vec{v}}\rangle\frac{d_2 f_2}{dt} + f_2\frac{d_2}{dt}\langle\dot{\vec{v}}\rangle = \frac{d_2}{dt}\int f_3\dot{\vec{v}}d^3\dot{v} = \int\dot{\vec{v}}\frac{d_2}{dt}f_3 d^3\dot{v}, \qquad (3.19)$$

or

$$f_2\frac{d_2}{dt}\langle\dot{\vec{v}}\rangle = \int\dot{\vec{v}}\frac{d_2}{dt}f_3 d^3\dot{v} - \langle\dot{\vec{v}}\rangle\frac{d_2}{dt}f_2 = \int\dot{\vec{v}}\frac{d_2}{dt}f_3 d^3\dot{v} - \langle\dot{\vec{v}}\rangle\frac{d_2}{dt}\int f_3 d^3\dot{v} =$$

$$= \int\dot{\vec{v}}\frac{d_2}{dt}f_3 d^3\dot{v} - \int\langle\dot{\vec{v}}\rangle\frac{d_2}{dt}f_3 d^3\dot{v} = \int(\dot{\vec{v}}-\langle\dot{\vec{v}}\rangle)\frac{d_2}{dt}f_3 d^3\dot{v} =$$

$$= \int(\dot{\vec{v}}-\langle\dot{\vec{v}}\rangle)\frac{\partial f_3}{\partial t}d^3\dot{v} + \int(\dot{\vec{v}}-\langle\dot{\vec{v}}\rangle)(\vec{v},\nabla_r f_3)d^3\dot{v} + \int(\dot{\vec{v}}-\langle\dot{\vec{v}}\rangle)(\langle\vec{v}\rangle,\nabla_v f_3)d^3\dot{v}. \qquad (3.20)$$

With the use of the third Vlasov equation (1.1)

$$\frac{\partial f_3}{\partial t} + \mathrm{div}_r\left[f_3\vec{v}\right] + \mathrm{div}_v\left[\dot{\vec{v}}f_3\right] + \mathrm{div}_{\dot{v}}\left[\langle\ddot{\vec{v}}\rangle f_3\right] = 0, \qquad (3.21)$$

expression (3.20) is of the form

$$f_2 \frac{d_2}{dt}\langle\dot{\vec{v}}\rangle = -\int(\dot{\vec{v}}-\langle\dot{\vec{v}}\rangle)\left(\text{div}_r[f_3\vec{v}]+\text{div}_v[\dot{\vec{v}}f_3]+\text{div}_{\dot{v}}[\langle\ddot{\vec{v}}\rangle f_3]\right)d^3\dot{v}+$$

$$+\int(\dot{\vec{v}}-\langle\dot{\vec{v}}\rangle)(\vec{v},\nabla_r f_3)d^3\dot{v}+\int(\dot{\vec{v}}-\langle\dot{\vec{v}}\rangle)(\langle\dot{\vec{v}}\rangle,\nabla_v f_3)d^3\dot{v} =$$

$$= -\int(\dot{\vec{v}}-\langle\dot{\vec{v}}\rangle)(\vec{v},\nabla_r f_3)d^3\dot{v} - \int(\dot{\vec{v}}-\langle\dot{\vec{v}}\rangle)(\vec{v},\nabla_v f_3)d^3\dot{v} - \int(\dot{\vec{v}}-\langle\dot{\vec{v}}\rangle)\text{div}_{\dot{v}}[\langle\ddot{\vec{v}}\rangle f_3]d^3\dot{v}+$$

$$+\int(\dot{\vec{v}}-\langle\dot{\vec{v}}\rangle)(\vec{v},\nabla_r f_3)d^3\dot{v}+\int(\dot{\vec{v}}-\langle\dot{\vec{v}}\rangle)(\langle\dot{\vec{v}}\rangle,\nabla_v f_3)d^3\dot{v} =$$

$$= -\int(\dot{\vec{v}}-\langle\dot{\vec{v}}\rangle)(\dot{\vec{v}}-\langle\dot{\vec{v}}\rangle,\nabla_v f_3)d^3\dot{v} - \int \dot{\vec{v}}\,\text{div}_{\dot{v}}[\langle\ddot{\vec{v}}\rangle f_3]d^3\dot{v} + \langle\dot{\vec{v}}\rangle\int\text{div}_{\dot{v}}[\langle\ddot{\vec{v}}\rangle f_3]d^3\dot{v} =$$

$$= -\int(\dot{\vec{v}}-\langle\dot{\vec{v}}\rangle)(\dot{\vec{v}}-\langle\dot{\vec{v}}\rangle,\nabla_v f_3)d^3\dot{v} + \int f_3\langle\ddot{\vec{v}}\rangle d^3\dot{v} = f_2\langle\langle\ddot{\vec{v}}\rangle\rangle - \int(\dot{\vec{v}}-\langle\dot{\vec{v}}\rangle)(\dot{\vec{v}}-\langle\dot{\vec{v}}\rangle,\nabla_v)f_3 d^3\dot{v},$$

that is

$$f_2\langle\langle\ddot{\vec{v}}\rangle\rangle = f_2\frac{d_2}{dt}\langle\dot{\vec{v}}\rangle + \int(\dot{\vec{v}}-\langle\dot{\vec{v}}\rangle)(\dot{\vec{v}}-\langle\dot{\vec{v}}\rangle,\nabla_v)f_3 d^3\dot{v},$$

which proves the third expression (3.10). The relations for the rest of the derivatives are obtained in a similar manner.

*Proof of Theorem 5*

The first expression in (3.22) coincides completely with the first expression in (3.10). Let us integrate the second expression from (3.10) over $\int d^3r$, we obtain

$$\int_{(\infty)} f_1\langle\langle\dot{\vec{v}}\rangle\rangle d^3r = N(t)\langle\langle\langle\dot{\vec{v}}\rangle\rangle\rangle = \int_{(\infty)} f_1\frac{d_1}{dt}\langle\vec{v}\rangle d^3r + \int_{(\infty)}\int_{(\infty)}(\vec{v}-\langle\vec{v}\rangle)(\vec{v}-\langle\vec{v}\rangle,\nabla_r f_2)d^3v d^3r. \quad (3.23)$$

In paper [4] we obtained the expression

$$\int_{(\infty)} f_1\frac{d_1}{dt}\langle\vec{v}\rangle d^3r = N(t)\langle\langle\langle\dot{\vec{v}}\rangle\rangle\rangle(t). \quad (3.24)$$

Using (3.24) in (3.23), we obtain

$$\int_{(\infty)}\int_{(\infty)}(\vec{v}-\langle\vec{v}\rangle)(\vec{v}-\langle\vec{v}\rangle,\nabla_r f_2)d^3v d^3r = 0. \quad (3.25)$$

Let us check the correctness of (3.25) by direct calculation.

$$I = \int_{(\infty)}\int_{(\infty)}(\vec{v}-\langle\vec{v}\rangle)(\vec{v}-\langle\vec{v}\rangle,\nabla_r f_2)d^3v d^3r =$$

$$= \int_{(\infty)}\vec{v}\left(\vec{v},\int_{(\infty)}\nabla_r f_2 d^3r\right)d^3v - \int_{(\infty)}\vec{v}d^3v\int_{(\infty)}(\langle\vec{v}\rangle,\nabla_r f_2)d^3r - \quad (3.26)$$

$$- \int_{(\infty)}\int_{(\infty)}\langle\vec{v}\rangle(\vec{v},\nabla_r f_2)d^3r d^3v + \int_{(\infty)}\int_{(\infty)}\langle\vec{v}\rangle(\langle\vec{v}\rangle,\nabla_r f_2)d^3r d^3v = I_1 - I_2 - I_3 + I_4,$$

where $I_i$, $i=1,..,4$ stand for the corresponding integrals in expression (3.26). Let us consider each integral separately.

$$I_1 = \int_{(\infty)} \vec{v}\left(\vec{v}, \int_{(\infty)} \nabla_r f_2 d^3r\right) d^3v = 0, \qquad (3.27)$$

as $\int_{(\infty)} \nabla_r f_2 d^3r = \int_{\Sigma_\infty} f_2 d\vec{\sigma} = 0$.

$$I_2 = \int_{(\infty)} \vec{v} d^3v \int_{(\infty)} (\langle\vec{v}\rangle, \nabla_r f_2) d^3r = \int_{(\infty)} (\langle\vec{v}\rangle, \nabla_r) \left(\int_{(\infty)} f_2 \vec{v} d^3v\right) d^3r = \int_{(\infty)} (\langle\vec{v}\rangle, \nabla_r)(f_1\langle\vec{v}\rangle) d^3r =$$
$$= \int_{(\infty)} \langle\vec{v}\rangle(\langle\vec{v}\rangle, \nabla_r f_1) d^3r + \int_{(\infty)} f_1(\langle\vec{v}\rangle, \nabla_r)\langle\vec{v}\rangle d^3r, \qquad (3.28)$$

$$I_3 = \int_{(\infty)}\int_{(\infty)} \langle\vec{v}\rangle(\vec{v}, \nabla_r f_2) d^3r d^3v = \int_{(\infty)}\int_{(\infty)} \langle\vec{v}\rangle \mathrm{div}_r[\vec{v} f_2] d^3r d^3v = \int_{(\infty)} \langle\vec{v}\rangle \mathrm{div}_r\left(\int_{(\infty)} [\vec{v} f_2] d^3v\right) d^3r =$$
$$= \int_{(\infty)} \langle\vec{v}\rangle \mathrm{div}_r\left[\langle\vec{v}\rangle f_1\right] d^3r = -\int f_1(\langle\vec{v}\rangle, \nabla_r)\langle\vec{v}\rangle d^3r,$$

a formula obtained in [22] is used here

$$\int \langle\vec{v}\rangle \mathrm{div}_r\left[\langle\vec{v}\rangle f_1\right] d^3r = -\int f_1(\langle\vec{v}\rangle, \nabla_r)\langle\vec{v}\rangle d^3r.$$

$$I_4 = \int_{(\infty)}\int_{(\infty)} \langle\vec{v}\rangle(\langle\vec{v}\rangle, \nabla_r) f_2 d^3r d^3v = \int_{(\infty)} \langle\vec{v}\rangle(\langle\vec{v}\rangle, \nabla_r)\left(\int_{(\infty)} f_2 d^3v\right) d^3r = \int_{(\infty)} \langle\vec{v}\rangle(\langle\vec{v}\rangle, \nabla_r f_1) d^3r. \qquad (3.29)$$

Using expressions (3.27)-(3.29) in (3.26), we obtain that $I=0$, that is the correctness of (3.25):

$$I = 0 - \int_{(\infty)} \langle\vec{v}\rangle(\langle\vec{v}\rangle, \nabla_r f_1) d^3r - \int_{(\infty)} f_1(\langle\vec{v}\rangle, \nabla_r)\langle\vec{v}\rangle d^3r + \int f_1(\langle\vec{v}\rangle, \nabla_r)\langle\vec{v}\rangle d^3r +$$
$$+ \int_{(\infty)} \langle\vec{v}\rangle(\langle\vec{v}\rangle, \nabla_r f_1) d^3r = 0,$$

which was to be proved. Thus, from (3.23), taking into account (3.25), it follows that expression (3.24) is true, that is

$$\langle\langle\langle\dot{\vec{v}}\rangle\rangle\rangle(t) = \left\langle\frac{d_1}{dt}\langle\vec{v}\rangle\right\rangle(t). \qquad (3.30)$$

Let us integrate the third expression in (3.10) over $\int d^3r d^3v$, we obtain

$$\int\limits_{(\infty)}\int\limits_{(\infty)} f_2 \left\langle\left\langle \ddot{\vec{v}} \right\rangle\right\rangle d^3 r d^3 v = N(t)\left\langle\left\langle\left\langle\left\langle \ddot{\vec{v}} \right\rangle\right\rangle\right\rangle\right\rangle = \int\limits_{(\infty)}\int\limits_{(\infty)} f_2 \frac{d_2}{dt}\left\langle \dot{\vec{v}} \right\rangle d^3 r d^3 v +$$
$$+ \int\limits_{(\infty)}\int\limits_{(\infty)}\int\limits_{(\infty)} \left(\dot{\vec{v}} - \left\langle \dot{\vec{v}} \right\rangle\right)\left(\dot{\vec{v}} - \left\langle \dot{\vec{v}} \right\rangle, \nabla_v\right) f_3 d^3 r d^3 v d^3 \dot{v}.$$
(3.31)

Let us consider the triple integral in expression (3.31)

$$I = \int\limits_{(\infty)}\int\limits_{(\infty)}\int\limits_{(\infty)} \left(\dot{\vec{v}} - \left\langle \dot{\vec{v}} \right\rangle\right)\left(\dot{\vec{v}} - \left\langle \dot{\vec{v}} \right\rangle, \nabla_v f_3\right) d^3 r d^3 v d^3 \dot{v} = \int\limits_{(\infty)}\int\limits_{(\infty)}\int\limits_{(\infty)} \dot{\vec{v}}\left(\dot{\vec{v}}, \nabla_v f_3\right) d^3 r d^3 v d^3 \dot{v} -$$
$$- \int\limits_{(\infty)}\int\limits_{(\infty)}\int\limits_{(\infty)} \dot{\vec{v}}\left(\left\langle \dot{\vec{v}} \right\rangle, \nabla_v f_3\right) d^3 r d^3 v d^3 \dot{v} - \int\limits_{(\infty)}\int\limits_{(\infty)}\int\limits_{(\infty)} \left\langle \dot{\vec{v}} \right\rangle\left(\dot{\vec{v}}, \nabla_v f_3\right) d^3 r d^3 v d^3 \dot{v} +$$
$$+ \int\limits_{(\infty)}\int\limits_{(\infty)}\int\limits_{(\infty)} \left\langle \dot{\vec{v}} \right\rangle\left(\left\langle \dot{\vec{v}} \right\rangle, \nabla_v f_3\right) d^3 r d^3 v d^3 \dot{v} = I_1 - I_2 - I_3 + I_4.$$
(3.32)

Let us transform each integral separately from (3.32)

$$I_1 = \int\limits_{(\infty)}\int\limits_{(\infty)}\int\limits_{(\infty)} \dot{\vec{v}}\left(\dot{\vec{v}}, \nabla_v f_3\right) d^3 r d^3 v d^3 \dot{v} = \int\limits_{(\infty)}\int\limits_{(\infty)} \left(\int\limits_{(\infty)} \nabla_v f_3 d^3 v\right) d^3 r d^3 \dot{v} = 0,$$
(3.33)

as $\int\limits_{(\infty)} \nabla_v f_3 d^3 v = 0$.

$$I_2 = \int\limits_{(\infty)}\int\limits_{(\infty)} \left(\left\langle \dot{\vec{v}} \right\rangle, \nabla_v\right)\left(\int\limits_{(\infty)} f_3 \dot{\vec{v}} d^3 \dot{v}\right) d^3 r d^3 v = \int\limits_{(\infty)}\int\limits_{(\infty)} \left(\left\langle \dot{\vec{v}} \right\rangle, \nabla_v\right)\left(f_2 \left\langle \dot{\vec{v}} \right\rangle\right) d^3 r d^3 v =$$
$$= \int\limits_{(\infty)}\int\limits_{(\infty)} \left\langle \dot{\vec{v}} \right\rangle\left(\left\langle \dot{\vec{v}} \right\rangle, \nabla_v f_2\right) d^3 r d^3 v + \int\limits_{(\infty)}\int\limits_{(\infty)} f_2 \left(\left\langle \dot{\vec{v}} \right\rangle, \nabla_v\right)\left\langle \dot{\vec{v}} \right\rangle d^3 r d^3 v,$$
(3.34)

$$I_3 = \int\limits_{(\infty)}\int\limits_{(\infty)} \left\langle \dot{\vec{v}} \right\rangle\left(\int\limits_{(\infty)} \left(\dot{\vec{v}}, \nabla_v\right) f_3 d^3 \dot{v}\right) d^3 r d^3 v = \int\limits_{(\infty)}\int\limits_{(\infty)} \left\langle \dot{\vec{v}} \right\rangle\left(\int\limits_{(\infty)} \operatorname{div}_v\left[f_3 \dot{\vec{v}}\right] d^3 \dot{v}\right) d^3 r d^3 v =$$
$$= \int\limits_{(\infty)}\int\limits_{(\infty)} \left\langle \dot{\vec{v}} \right\rangle \operatorname{div}_v\left(\int\limits_{(\infty)}\left[f_3 \dot{\vec{v}}\right] d^3 \dot{v}\right) d^3 r d^3 v = \int\limits_{(\infty)}\int\limits_{(\infty)} \left\langle \dot{\vec{v}} \right\rangle \operatorname{div}_v\left[f_2 \left\langle \dot{\vec{v}} \right\rangle\right] d^3 r d^3 v =$$
$$= -\int\limits_{(\infty)}\int\limits_{(\infty)} f_2 \left(\left\langle \dot{\vec{v}} \right\rangle, \nabla_v\right)\left\langle \dot{\vec{v}} \right\rangle d^3 r d^3 v$$
(3.35)

$$I_4 = \int\limits_{(\infty)}\int\limits_{(\infty)} \left\langle \dot{\vec{v}} \right\rangle\left(\left\langle \dot{\vec{v}} \right\rangle, \nabla_v\right)\left(\int\limits_{(\infty)} f_3 d^3 \dot{v}\right) d^3 r d^3 v = \int\limits_{(\infty)}\int\limits_{(\infty)} \left\langle \dot{\vec{v}} \right\rangle\left(\left\langle \dot{\vec{v}} \right\rangle, \nabla_v f_2\right) d^3 r d^3 v.$$
(3.36)

Using expressions (3.33)-(3.36) in (3.32), we obtain

$$I = 0 - \int\limits_{(\infty)}\int\limits_{(\infty)} \left\langle \dot{\vec{v}} \right\rangle\left(\left\langle \dot{\vec{v}} \right\rangle, \nabla_v f_2\right) d^3 r d^3 v - \int\limits_{(\infty)}\int\limits_{(\infty)} f_2 \left(\left\langle \dot{\vec{v}} \right\rangle, \nabla_v\right)\left\langle \dot{\vec{v}} \right\rangle d^3 r d^3 v +$$
$$+ \int\limits_{(\infty)}\int\limits_{(\infty)} f_2 \left(\left\langle \dot{\vec{v}} \right\rangle, \nabla_v\right)\left\langle \dot{\vec{v}} \right\rangle d^3 r d^3 v + \int\limits_{(\infty)}\int\limits_{(\infty)} \left\langle \dot{\vec{v}} \right\rangle\left(\left\langle \dot{\vec{v}} \right\rangle, \nabla_v f_2\right) d^3 r d^3 v = 0.$$
(3.37)

Considering (3.37), expression (3.31) is of the form

$$\left\langle\left\langle\left\langle\left\langle \ddot{\vec{v}} \right\rangle\right\rangle\right\rangle\right\rangle(t) = \left\langle\left\langle \frac{d_2}{dt}\left\langle \dot{\vec{v}} \right\rangle\right\rangle\right\rangle(t). \tag{3.38}$$

One may obtain relations for the derivatives of a higher order in a similar manner.

## Proof of Theorem 6

Let us write the second equation in (3.10) in the integral form

$$f_1\left\langle\left\langle \dot{\vec{v}} \right\rangle\right\rangle(\vec{r},t) = f_1 \frac{d_1}{dt}\left\langle \vec{v} \right\rangle(\vec{r},t) + \int\limits_{(\infty)} (\vec{v}-\left\langle\vec{v}\right\rangle)(\vec{v}-\left\langle\vec{v}\right\rangle, \nabla_r f_2)d^3v,$$

$$\int\limits_{(\infty)} f_2 \left\langle \dot{\vec{v}} \right\rangle d^3v = \int\limits_{(\infty)} f_2 \frac{d_1}{dt}\left\langle \vec{v} \right\rangle d^3v + \int\limits_{(\infty)} (\vec{v}-\left\langle\vec{v}\right\rangle)(\vec{v}-\left\langle\vec{v}\right\rangle, \nabla_r f_2)d^3v, \tag{4.4}$$

from this it follows

$$f_2\left\langle \dot{\vec{v}} \right\rangle = f_2 \frac{d_1}{dt}\left\langle \vec{v} \right\rangle + (\vec{v}-\left\langle\vec{v}\right\rangle)(\vec{v}-\left\langle\vec{v}\right\rangle, \nabla_r f_2) + f_2\vec{\eta}, \tag{4.5}$$

or

$$\left\langle \dot{\vec{v}} \right\rangle = \frac{d_1}{dt}\left\langle \vec{v} \right\rangle + (\vec{v}-\left\langle\vec{v}\right\rangle)(\vec{v}-\left\langle\vec{v}\right\rangle, \nabla_r S_2) + \vec{\eta},$$

where $\vec{\eta}$ − some vector-function having property (4.3), which was to be proved.

## Proof of Theorem 7

Let condition (4.11) to be fulfilled, then from (4.6) (theorem 6) it follows that

$$\left\langle \dot{\vec{v}} \right\rangle = \frac{d}{dt}\left\langle \vec{v} \right\rangle = \frac{\vec{F}}{m}, \tag{4.14}$$

and from (4.8) it follows that

$$Q_2 = \mathrm{div}_v \left\langle \dot{\vec{v}} \right\rangle = 0. \tag{4.15}$$

From (4.14) and (4.15) it follows that the Vlasov equation (4.1) is of the form

$$\Pi_2 S_2 = 0, \tag{4.16}$$

or

$$\frac{\partial f_2}{\partial t} + (\vec{v}, \nabla_r f_2) + \left(\frac{\vec{F}}{m}, \nabla_v f_2\right) = 0,$$

which corresponds to (4.9) at $\Lambda = 0$. At a constant number of particles $N = const$ according to (2.22) (Theorem 2) and (4.16), entropy $H_2$ satisfies the equation

$$\frac{dH_2}{dt} = \langle \Pi_2 S_2 \rangle = 0, \qquad (4.17)$$

That is $H_2 = const$, which was to be proved.